# The Case for a DNANF 1Pb/s Trans-Atlantic Submarine Cable

Pierluigi Poggiolini[(1)], Francesco Poletti[(2)]

[(1)] DET, Politecnico di Torino, C.so Duca Abruzzi 24, 10129, Torino, Italy, pierluigi.poggiolini@polito.it
[(2)] ORC, Univ. of Southampton, UK, and Microsoft Azure Fiber, The Quadrangle, Romsey UK, frap@orc.soton.ac.uk

**Abstract** *The recent progress in low-loss hollow-core fibers allows to speculate on the possibility of building a transatlantic submarine cable that can achieve the goal of 1 Pb/s per direction, leveraging bidirectional transmission, and at the same time drastically increase span length, theoretically to 200km. In this version, we add the analysis of the impact of Rayleigh backscattering.* ©2025 The Author(s)

**Introduction**

Recent results on hollow-core fibers of the DNANF type have shown loss below 0.11 dB/km [1][2], with hints that some DNANF segments had much lower loss in the C-band, on the order of 0.05-0.08 dB/km. This is compatible with theory, that shows that loss values as low as 0.05 dB/km should be indeed possible [12]. At the same time, bidirectional transmission has also been recently demonstrated, leveraging the ultra-low backscattering of DNANFs [3]. In addition, the possibility of ultra-long system spans with DNANF, reaching up to 150-250km thanks to the combination of their low-loss and ultra-low non-linearity, was pointed out in [4].

All these circumstances, specific to DNANF, make it possible to hypothesize building a trans-atlantic (TAT) cable whose total throughput reaches the 1 Pb/s threshold, while allowing to stretch the span length to unprecedented values, on the order of 200km. This latter aspect would allow important savings and would also greatly ease the power supply hurdle which is one of the key factors preventing submarine cables from scaling up throughput.

This paper analyses this scenario, making the case that the potential for DNANFs to allow TAT cables to be realized with such features is theoretically there, and that it is worth further and deeper investigation, as well as experimental exploration.

Since the system is bi-directional, Rayleigh backscattering (RBS) may play a role. **The original version of this paper, published at ECOC 2025, was written under the assumption that Rayleigh backscattering from DNANF was so low that it would be negligible vs. other sources of disturbance. In this follow-up, we add an Appendix in which we perform a RBS analysis and explicitly calculate its impact. The analysis confirms that RBS has little or negligible effect on the scenarios addressed here.**

**The 1 Pb/s DNANF system configuration**
The cable is assumed to be a standard-diameter submarine cable, containing 26 DNANFs in total. This number of fibers is a conjecture based on the following reasoning. New cables (such as Anjana [5] or Medusa [6]) carry 24 solid-core fiber (SCF) pairs, i.e., 48 SCFs. Such SCFs are reduced-diameter ones (conceivably 200μm). DNANFs necessarily have larger diameter so fewer would fit a similarly built cable. However, through simple math, the same cable cross-sec-tion fitting 48 200μm SCFs would presumably fit 26 DNANFs with diameter 270 μm, a value that we assume to be sufficient to support DNANF.

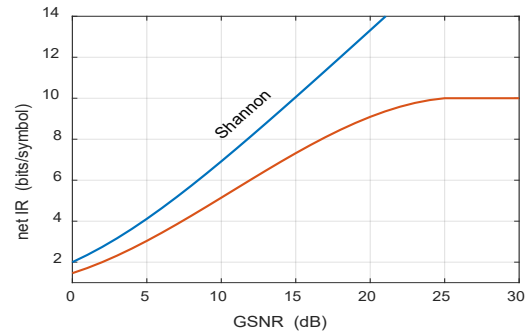

**Fig. 1:** Transceiver net information rate vs. GSNR.

We then assume that traffic is carried in the C-band only, over a bandwidth of 5THz. The DNANFs are used bi-directionally so 26 full C-band combs propagate in each direction. We consider state-of-the-art transceivers with probabilistic constellation shaping (PCS). We adopt for such TXRs the (non-model-specific) net-information-rate vs. GSNR curve shown in Fig.1. We assume that the channel symbol rate is 73.5 GBaud and the channel spacing is 75 GHz, as for the ultra-long-haul PSC experiment n.2 in [7].

The DNANF is assumed to have dispersion 3ps/(nm km) and non-linearity coefficient $5 \cdot 10^{-4}$ 1/(W km). An inter-modal-interference (IMI) of -65 dB/km is considered.

At each repeater a DNANF-to-SCF transition is needed. These components have been reported with loss lower than 0.2dB [13]. We assume here a conservative value of 0.7dB. We also assume negligible back-reflection. Next, a circulator is needed to extract the incoming signal and insert the outgoing one for the counter-propagating direction. C-band circulators are commercially available with loss <1dB. We assume 1dB. We then add another 0.3dB loss margin so that loss before the EDFA input is 2dB. The EDFA noise figure is set to 4.6 dB [7]. At the output of the EDFA the signal encounters another circulator and a SCF-to-DNANF transition. We assume the same conservative loss as at the EDFA input (2dB), including a 0.3dB margin.

Finally, we set the span length to the challenging value of 200km. The total cable length is assumed to be 6600km, similar to the MAREA TAT cable [8].

**Performance results**
The analysis method is the same as described in [9],[4]. The resulting GNSR accounts for ASE noise, IMI and NLI noise, the latter through the GN-model. TRX noise is embedded in the curve Fig.1, which is used to obtain the net throughput from GSNR. The droop effect [10] is neglected since the operating GSNRs (see later) are large.

In Fig.2a and Fig.2b the contour plots of the GSNR and of the net overall cable throughput are shown, respectively. The abscissa is DNANF loss and the ordinate is the total output power per EDFA. Note that the power entering each DNANF is 2dB lower than shown in Fig.2, due to the circulator and SCF-DNANF transition present at the EDFA output and whose combined loss, as mentioned, was assumed to be 2dB.

Fig.2b shows that 1Pb/s, currently considered as a landmark achievement for possible future cables, is ideally reachable at 0.07dB/km DNANF loss, with a C-band EDFA output power of 22.5 dBm. It is also reachable at about 0.06 and 0.05 dB/km loss, at EDFA output powers of 20.3 and 18 dBm, respectively. It should also be noted that a still remarkable 0.9 Pb/s is possible, at the mentioned DNANF loss values, with a reduction of 2dB EDFA power. All these operating points have a GSNR greater than 14dB (see Fig.2a), which ensures that the droop effect is negligible [10].

In the following, we consider as "reference" the operating point: 0.06dB/km loss, 20.3dBm per EDFA, 1Pb/s. While very challenging, a loss of 0.06dB/km over the C-band is compatible with theoretical predictions and some of the experimental results [1],[2]. Also, while the EDFA output power of 20.3 dBm is somewhat large for a submarine system, it would be compatible with the cable power supply limit currently considered to be about 18kW [7]. The reason is that the number of submerged repeaters is only 32, as opposed to 80-100 for SCF transatlantic cables [7]. Power dissipation could be estimated at 6.6kW from the cable (assuming 1A current and 1$\Omega$/km cable resistance) and 5.76kW from repeaters. The latter number assumes 180W per repeater, a figure compatible with the data shown in [11]. The total DNANF system power consumption would then be less than 13kW, well below the 18kW limit.

In fact, the DNANF TAT cable at the "reference" operating point is not power-limited, but space-limited. It appears that 18kW would be enough to support substantially more than the assumed 26 DNANFs (about 40-50). This suggests that a technology research path towards bigger cables, capable of carrying more DNANFs, could potentially get close to 2Pb/s per TAT cable within today's power supply limits.

Finally, we explored different span lengths

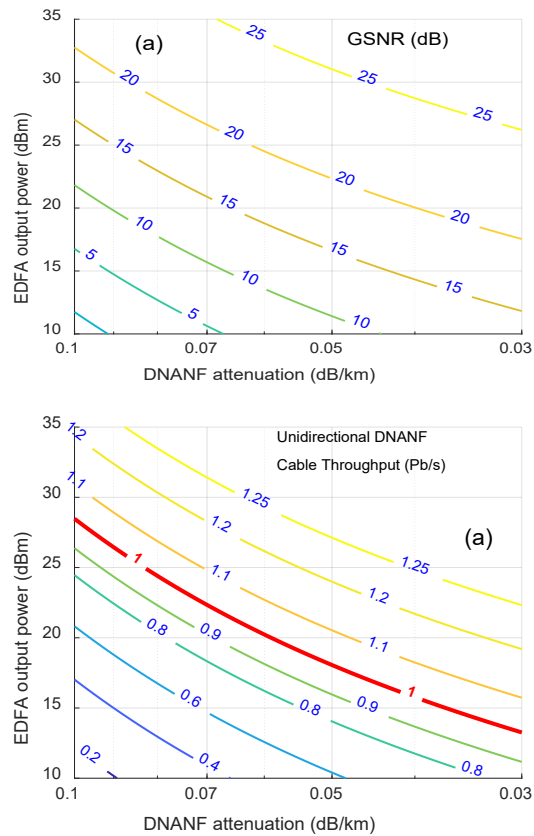

**Fig. 2: 6600km link with 200km spans, 26 DNANF.** Contour plots, vs. DNANF attenuation and vs. EDFA output power of **(a):** the GSNR in dB and **(b):** the total throughput of the DNANF cable in one direction.

than 200km. The required output power per EDFA to achieve 1 Pb/s net throughput is shown in Fig.3. The plot shows that at 0.06dB/km loss, reducing the span length to 170km may reduce the needed EDFA output power by about 1dB. Conversely, 220-230km span length would be attainable if it was 1dB higher. Similar curves for 0.05 and 0.07dB/km loss are also shown

**Discussion and competing technologies**
A 1Pb/s throughput DNANF TAT cable would be about double the throughput of the very recent SCF Anjana cable, which completed deployment in Oct. 2024. But one of the most remarkable features of the DNANF cable would be its small number of repeaters (32).

Many recent studies have focused on reduced-diameter and multi-core fibers. However, the key insurmountable difference between any system based on DNANF vs. systems based on any type of single or multicore SCF is indeed the span length. The typical 60 to 70km length span length in SCF transatlantic systems cannot be exceeded because of the fundamental loss and non-linearity features of SCFs. The possibility for DNANF cables to aim at 3x the span length, and perhaps 4x if loss reaches down to 0.05dB/km, sets them completely apart from any SCFs system. Besides such prominent advantages, the use of DNANFs would bring about further welcome consequences. One is the drastic mitigation of EEPN, since DNANF dispersion is about 1/8 that of PSCF. For the same reason, DSP complexity would shrink substantially: while not a deciding factor, it is certainly a welcome plus.

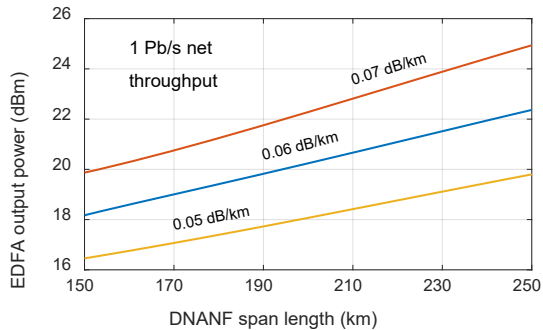

**Fig. 3:** EDFA output power vs. span length in the transatlantic DNANF cable, to achieve 1 Pb/s total net throughput per direction, for three different values of the DNANF loss.

Finally, another key aspect is latency. Transatlantic propagation takes 33ms in SCF and 22ms in DNANF. While for most users the difference may not be meaningful, for some users this difference is very valuable. In this respect too, of course, there is no possibility for SCF to compete.

Though this paper focuses on transatlantic links, the advantages described above broadly apply to transpacific cables as well, which we do not address here for lack of space.

### Challenges

Several challenges stand in the way of realizing the DNANF cable systems described above.

One is mass production. Educated guesses may place total DNANF production in 2024 at a few thousands km. A single TAT cable carrying 26 DNANFs would require about 170,000km. The gap between these figures is evident. Also, the remarkable low-loss results [1],[2] were shown over short stretches: achieving 0.06-0.07 dB/km loss consistently over multi-thousand km production fiber is a much more challenging feat.

Another aspect that needs attention is the ability for DNANF to reliably achieve low values of IMI. We assumed -65dB/km. A value of $-60$ dB/km would still be not too critical as it would lead to requiring 21.2dB EDFA output power to achieve 1 Pb/s at 0.06 dB/km loss, up 0.9 dB with respect to Fig.2b. However, above $-60$dB/km performance degrades steeply so it is indispensable that IMI be kept under control. On the other hand, very recent results hint at some DNANF achieving -70dB/km, which is reassuring [14].

Though remarkable experiments have confirmed the bi-directional capability of DNANF [3], still the reliability of enabling components such as circulators needs to be confirmed for submarine use. But perhaps the most critical component in the chain is the DNANF-to-SCF transition. This component needs to have loss below 1dB and negligible back-reflection. Very encouraging results have appeared in the literature [13], but further development and ruggedization is needed.

Also, we have assumed in our calculations that DNANF can be mass-produced with cutting-edge performance at an outer diameter of only 270 μm. This is a bold assumption, since the air-filled cavity of DNANF has a very wide diameter of approximately 90μm. On the other hand, this requirement might relax in the future, since it is ideally possible that bigger cables capable of hosting 26 or even more DNANFs at, say, 280-300μm diameter will be manufactured, once the advantage of using DNANF in the transoceanic realm is definitively proven.

Another known problem is the presence of gas contaminants in the hollow core, namely water and carbon dioxide, the latter causing thin absorption lines to show up in C and L band. Contaminants must be eliminated and perhaps transmission techniques tolerating some residual level should be developed.

Finally, an overarching challenge is obviously cost. DNANF cost must come down to the level needed to make DNANF transoceanic cables economically competitive. This aspect is well beyond the scope of this paper and has to do with production technology development as well as techno-economics at various levels, but it is of course the bottom-line enabler for DNANF TAT cables to become a reality.

### Conclusion

The recent progress in low-loss hollow-core fibers allowed us to speculate on the possibility of building a transatlantic submarine cable that can achieve the iconic goal of 1 Pb/s per direction and, at the same time, drastically increase span length, to a previously unthinkable 200km. Based on recent results, both theoretical and experimental, and reasonable hypotheses, these systems appear to be in the realm of potentially possible achievements.

A number of challenges stand in the way, some technological, some economical, of which we tried to provide a partial list. This said, the unique features of DNANF make the case of a DNANF submarine cable extremely compelling and, in our opinion, worth intense investigation.

**Acknowledgements**

This work was partially supported by the PhotoNext Center of Politecnico di Torino, and the European Union under the Italian National Recovery and Resilience Plan (NRRP) of NextGenerationEU, partnership on "Telecommunications of the Future" (PE00000001 - program "RESTART").

**Appendix: Rayleigh Backscattering Analysis**
This Appendix derives the equation for the total Rayleigh backscattered power in a multi-span bi-directional optical fiber link with identical spans, under the assumptions of transparency (loss exactly compensated for by amplifiers in both forward and backward directions at each span) and uniform fiber properties.

Regarding notation:
- $P_{\text{RBS}}$: Rayleigh backscattered power
- $P_{\text{ch}}(z)$: channel power at the coordinate $z$ along a span; as a shorthand, the writing $P_{\text{ch}}$ without the argument will indicate $P_{\text{ch}} = P_{\text{ch}}(0)$, that is the launch power into each span
- $B$: backscattering coefficient (in $\text{km}^{-1}$)
- $\alpha$: *power* attenuation coefficient (in $\text{km}^{-1}$)
- $L_{\text{span}}$: span length (in km)
- $N_{\text{span}}$: number of spans

*Assumptions*
The derivation relies on the following assumptions:
- the link consists of $N_{\text{span}}$ identical spans, each with length $L_{\text{span}}$ and uniform properties (in particular uniform power attenuation $\alpha$ and uniform backscattering coefficient $B$).
- inline optical amplifiers are placed at the end of each span, providing gain:

$$G = \exp(\alpha L_{\text{span}})$$

to compensate exactly for the span loss *in both forward and backward directions*, ensuring transparency (net gain of 1 per span).
- Consequently, forward signal power is reset to the launch value $P_{\text{ch}} = P_{\text{ch}}(0)$ at the start of each span due to amplification. The same is done in the backward direction.
- Secondary effects like double Rayleigh backscattering are neglected.

*Single Span Backscattered Power*
The differential backscattered power from a small segment $dz$ at distance $z$ within a span (where $z = 0$ is the start of the span) is:

$$dP_{\text{RBS}} = P_{\text{ch}}(z) B dz$$

where $P_{\text{ch}}(z)$ is the forward-propagating power at $z$.

The forward power attenuates as:
$$P_{\text{ch}}(z) = P_{\text{ch}} \exp(-\alpha z)$$
Thus:
$$dP_{\text{RBS}} = P_{\text{ch}} \exp(-\alpha z) B dz.$$

The Rayleigh backscattered power travels back to the start of the span, undergoing additional attenuation over a length equal to $z$:
$$dP_{\text{RBS}} = P_{\text{ch}} B \exp(-2\alpha z) dz.$$
Integrating over the span length $L_{\text{span}}$:
$$P_{\text{RBS}} = \int_0^{L_{\text{span}}} P_{\text{ch}} B \exp(-2\alpha z) dz$$
$$= P_{\text{ch}} B \cdot \frac{1}{2\alpha} \left[ 1 - \exp(-2\alpha L_{\text{span}}) \right].$$

*Effect of Amplification in a Single Span*
In a transparent *bidirectional* system, the backscattered power from within the span is amplified by a gain $G = \exp(\alpha L_{\text{span}})$ when reaching the start of the span traveling backward. This amplification applies to the entire backscattered power from the span, effectively boosting it by $\exp(\alpha L_{\text{span}})$:

$$P_{\text{RBS}} = P_{\text{ch}} B \cdot \frac{1}{2\alpha} \left[ 1 - \exp(-2\alpha L_{\text{span}}) \right] \cdot$$
$$\cdot \exp(\alpha L_{\text{span}})$$

*Multi-Span Extension*
For $N_{\text{span}}$ spans, the backscattered power from each span contributes equally to the total at the input, as the return path through previous spans is transparent (net gain 1 per span due to

balanced loss and amplification). Therefore:

$$P_{\text{RBS}} = N_{\text{span}} P_{\text{ch}} B \cdot \frac{1}{2\alpha} \cdot$$
$$\cdot \left[ 1 - \exp(-2\alpha L_{\text{span}}) \right] \cdot \exp(\alpha L_{\text{span}})$$

This equation quantifies the cumulative Rayleigh backscattered power in a multi-span transparent link.

Interestingly, the above equation can be simplified as:

$$P_{\text{RBS}} = N_{\text{span}} P_{\text{ch}} B \cdot \frac{\sinh(\alpha L_{\text{span}})}{\alpha}$$
$$= N_{\text{span}} L_{\text{span}} P_{\text{ch}} B \cdot \frac{\sinh(\alpha L_{\text{span}})}{\alpha L_{\text{span}}}$$
$$= N_{\text{span}} L_{\text{span}} P_{\text{ch}} B \cdot \sinhc(\alpha L_{\text{span}})$$

where sinhc(x)=sinh(x)/x is called the hyperbolic sinc function. It is then interesting to notice that $\alpha L_{\text{span}}$ relates to the span loss. So, the formula can be rewritten using quantities that are more familiar. Using the total span loss in dB $A_{\text{dB}}^{\text{span}}$ then:

$$P_{\text{RBS}} = L_{\text{tot}} P_{\text{ch}} B \cdot \sinhc \left( \frac{A_{\text{dB}}}{10 \cdot \log_{10}(e)} \right)$$
$$= L_{\text{tot}} P_{\text{ch}} B \cdot \sinhc \left( \frac{\alpha_{\text{dB}} \cdot L_{\text{span}}}{10 \cdot \log_{10}(e)} \right)$$

For an ideal system with lossless fiber (or ideal distributed amplification), then:

$$P_{\text{RBS}} = L_{\text{tot}} P_{\text{ch}} B$$

If lumped amplification is used instead to compensate for loss, then a "enhancing factor"

$$\sinhc \left( \frac{A_{\text{dB}}}{10 \cdot \log_{10}(e)} \right) > 1$$

kicks in. We call it "enhancing factor" because its minimum value is 1, for ideal distributed amplification, that is for $A_{\text{dB}} = 0$. Any argument of the sinhc greater than 0 causes it to be greater than 1 and thus enhances RBS.

So, we define the "enhancing factor":

$$\text{enh}(A_{\text{dB}}) = \sinhc \left( \frac{A_{\text{dB}}}{10 \cdot \log_{10}(e)} \right)$$

and the RBS equation then becomes:

$$P_{\text{RBS}} = L_{\text{tot}} \cdot P_{\text{ch}} \cdot B \cdot \text{enh}(A_{\text{dB}})$$

*System impact*
The RBS noise should be added to the other sources of noise, like ASE, NLI or IMI. It is also possible to gauge its impact by looking at the GSNR due to RBS alone. It is easily found by re-arranging the above formula:

$$\text{GSNR}_{\text{RBS}} = \frac{P_{\text{ch}}}{P_{\text{RBS}}} = \frac{1}{L_{\text{tot}} \cdot B \cdot \text{enh}(A_{\text{dB}})}$$

Note the interesting circumstance that this GSNR is independent of launch power.

*Calculating RBS for the 1 Pb/s system*
The relevant parameters for the scenarios considered in this paper are:

- $L_{\text{tot}}$ = 6600 km
- $L_{\text{span}}$ = 200 km
- we consider DNANF loss equal to 0.05 dB/km or 0.07 dB/km, resulting in 10 dB or 14 dB respectively
- $B$: according to direct measurements carried out in [15], a typical value for NANF is -70 dB/km.

Using these numbers, we get that
- $\text{GSNR}_{\text{RBS}} = 28.48$ **dB for DNANF loss of 0.05 dB/km**
- $\text{GSNR}_{\text{RBS}} = 25.9$ **dB for DNANF loss of 0.07 dB/km**

These values of GSNR are small enough to cause almost no change in Fig.2 on the 1 Pb/s line (solid red). At 0.05 dB/km there is a slight up-shift in the sense that the required launch power to achieve 1 Pbit/s is 0.3 dB. At 0.07 dB/km it grows to 0.5 dB.

Note that if the span length is reduced, the impact is also reduced. Overall, the impact of Rayleigh is modest to negligible, in the considered cases.

### References

[1] Chen, Y., Petrovich, M. N., Numkam Fokoua, E., Adamu, A. I., Hassan, M. R. A., Sakr, H., Slavík, R., Gorajoobi, S. B., Alonso, M., Ando, R. F., Papadimopoulos, A., Varghese, T., Wu, D., Ando, M. F., Wisniowski, K., Sandoghchi, S. R., Jasion, G. T., Richardson, D. J., & Poletti, F. "Hollow Core DNANF Optical Fiber with <0.11 dB/km Loss," In Optical Fiber Communication Conference (OFC) 2024, PD paper Th4A.8.F, San Diego (CA), March 2024. https://doi.org/10.1364/OFC.2024.Th4A.8

[2] Y. Xiong, D. Zhang, S. Gao, D. Ge, Y. Sun, R. Zhao, Y. Xiao, Z. Yang, D. Wang, H. Li, X. Duan, W. Ding, and Y. Wang, "Field-deployed hollow-core fibre cable with 0.11 dB/km loss," in Proc. Eur. Conf. Opt. Commun. (ECOC), Post-deadline Paper Th3B.8, Frankfurt (D), 2024.

[3] L. Feng, A. Zhang, J. Luo, L. Zhang, P. Li, J. Chu, Y. Liu, X. Huo, J. Li, and C. Zhang, "Demonstration of single-span 100 km hollow core fiber bidirectional transmission with 1 Tb/s/λ real-time signals," in Proc. Opt. Fiber Commun. Conf. (OFC), San Francisco, CA, USA, Apr. 2025.

[4] P. Poggiolini, G. Bosco, Y. Jiang and F. Poletti, "The Potential for Span Length Increase with NANF," 2023 IEEE Photonics Conference (IPC), Orlando, FL, USA, 2023. doi: 10.1109/IPC57732.2023.10360569.

[5] https://www.submarinenetworks.com/en/systems/trans-atlantic/anjana?utm_source=chatgpt.com

[6] https://www.submarinenetworks.com/en/systems/asia-europe-africa/medusa?utm_source=chatgpt.com

[7] A. C. Meseguer et al., "Exploring the Potential of Multicore Fibers for Submarine SDM Systems: Experimental Validation and Powering Constraints," in IEEE/OPTICA Journal of Lightwave Technology, early access, March 2025. doi: 10.1109/JLT.2025.3555853.

[8] https://www.submarinenetworks.com/en/systems/trans-atlantic/marea?utm_source=chatgpt.com

[9] P. Poggiolini and F. Poletti, "Opportunities and Challenges for Long-Distance Transmission in Hollow-Core Fibres," in IEEE/OPTICA Journal of Lightwave Technology, vol. 40, no. 6, pp. 1605-1616, 15 March15, 2022, doi: 10.1109/JLT.2021.3140114.

[10] A. Bononi, J. -C. Antona, M. Lonardi, A. Carbo-Méseguer and P. Serena, "The Generalized Droop Formula for Low Signal to Noise Ratio Optical Links," in *Journal of Lightwave Technology*, vol. 38, no. 8, pp. 2201-2213, 15 April15, 2020, doi: 10.1109/JLT.2020.2966145.

[11] A. Carbó Meseguer et al., "On the Road to 1-Pbps systems: Experimental Demonstration of an Energy Efficient 500-Tbps Transatlantic Cable with 200-μm Outer Diameter Fibers," 2023 Optical Fiber Communications Conference and Exhibition (OFC), San Diego, CA, USA, 2023. doi: 10.1364/OFC.2023.Tu2G.5.

[12] N. Fokoua, S. A. Mousavi, G. T. Jasion, D. J. Richardson, and F. Poletti, "Loss in hollow-core optical fibers: mechanisms, scaling rules, and limits," OPTICA Adv. Opt. Photon., review paper, vol. 15, pp. 1–85, 2023. https://doi.org/10.1364/AOP.470592

[13] A. Zhong et al., "Connecting Hollow-Core and Standard Single-Mode Fibers With Perfect Mode-Field Size Adaptation," in Journal of Lightwave Technology, vol. 42, no. 6, pp. 2124-2130, 15 March 15, 2024, doi: 10.1109/JLT.2023.3329738.

[14] Marco Petrovich, et al., "First broadband optical fibre with an attenuation lower than 0.1 decibel per kilometre," arXiv, March 27[th] 2025. https://doi.org/10.48550/arXiv.2503.21467

[15] Radan Slavík, et al "Optical time-domain backscattering of antiresonant hollowcore fibers," Opt.Express 30,31310-31321(2022)